\documentclass[aps,pra,reprint,superscriptaddress,nofootinbib]{revtex4-2}

\usepackage{amsmath,amssymb,amsthm,bm,bbm}
\usepackage{graphicx}
\usepackage{xcolor}
\usepackage{hyperref}
\hypersetup{colorlinks=true, citecolor=blue, linkcolor=blue, urlcolor=blue}
\usepackage{tikz}
\usepackage{braket}
\usepackage{float}

\begin{document}

\title{Planted-Solution Pauli Hamiltonians as a Quantum Benchmarking Primitive}

\author{Amir Kalev}
\email{amirk@isi.edu}
\affiliation{Information Sciences Institute, University of Southern California, Arlington, VA 22203, USA}
\author{Itay Hen}
\email{itayhen@isi.edu}
\affiliation{Information Sciences Institute, University of Southern California, Marina del Rey, CA 90292, USA}
\affiliation{Department of Physics and Astronomy, and Center for Quantum Information Science \& Technology, University of Southern California, Los Angeles, California 90089, USA}

\date{\today}

\begin{abstract}
\noindent 
We introduce a construction of Pauli Hamiltonians with exactly known ground-state energies, intended as reference instances for ground-state energy estimation algorithms. The construction embeds a planted block-product state as the simultaneous ground state of a sum of frustration-free local clauses on overlapping supports, exposes the resulting model only as a polynomial-size linear combination of Pauli operators, and admits optional Clifford conjugation that preserves the spectrum. The framework subsumes classical planted constraint-satisfaction problems as a diagonal special case, providing a direct embedding channel through which classical hardness properties can be inherited. 
Open-source software, certification keys, and example instances are made publicly available.
\end{abstract}

\maketitle

\section{Introduction}
The ability to compute ground-state properties of quantum many-body Hamiltonians lies at the heart of quantum simulation, condensed-matter physics, quantum chemistry, and combinatorial optimization~\cite{feynman1982simulating,lloyd1996universal,georgescu2014quantum}. Determining ground-state energies, identifying ground states, and predicting dynamical observables are tasks that are widely believed to be classically intractable in the worst case~\cite{kitaev2002classical,kempe2006complexity}, and they therefore constitute prime targets for quantum advantage. As quantum hardware and quantum algorithms continue to mature~\cite{preskill2018quantum}, a central question emerges: how can we reliably benchmark quantum devices and algorithms that are designed to solve problems beyond the reach of classical computation?

Benchmarking quantum computers presents a difficulty that goes beyond ordinary performance evaluation. For classically tractable problems, correctness can be verified by comparison with trusted classical reference calculations. However, for problems that are intended to demonstrate quantum advantage, classical verification is precisely what is unavailable. In this regime, it is often unclear whether a reported quantum result is correct, approximately correct, or entirely spurious. This verification gap poses a major obstacle to the systematic assessment of quantum algorithms and hardware, particularly for ground-state and dynamical simulation tasks~\cite{eisert2020quantum}.

This tension has motivated the development of benchmarking strategies that interpolate between tractability and hardness. Ideally, benchmark instances should be sufficiently complex to stress both quantum and classical solvers, while still admitting exact reference solutions against which results can be objectively evaluated. At the same time, these instances should avoid visible algebraic or structural features that allow algorithms to bypass the intended difficulty. Designing such benchmarks in a controlled and scalable manner remains an open challenge, especially for analog and near-term quantum devices~\cite{boixo2018characterizing,arute2019quantum}.

A powerful paradigm for addressing this challenge originates in the theory of planted solutions. In the classical setting, planted-solution constructions were introduced in the study of satisfiability (SAT) and constraint satisfaction problems as a means of generating hard-looking instances with known solutions~\cite{jerrum1992large,achlioptas2001two}. By embedding a hidden satisfying assignment into an otherwise random-looking formula, one obtains instances that are efficiently verifiable yet conjectured to be difficult to solve without access to the planted structure. Planted SAT instances have played a central role in average-case complexity theory, algorithmic phase transitions, and the empirical evaluation of heuristic solvers~\cite{barthel2002hiding,krzakala2007gibbs,achlioptas2006solution}. Importantly, not all planted ensembles are hard: the difficulty depends on the specific distribution of clauses and the structure of the planted solution. The value of the planted paradigm lies in its ability to produce \emph{controlled} instances with known solutions, spanning a range from spectrally benign to conjectured-hard, thereby enabling systematic benchmarking.

The planted-solution idea was subsequently adapted to optimization problems and spin-glass models, where optimal configurations or ground states are embedded by construction~\cite{hen2015probing}. Such instances have been widely used to benchmark classical heuristics and quantum annealers, enabling controlled studies of scaling behavior, energy landscape structure, and algorithmic dynamics~\cite{boixo2014evidence,Hen2019EquationPlanting,Kowalsky2022_3R3X}. 

In the context of quantum Hamiltonians, benchmarking becomes more subtle. Many Hamiltonians with known ground states belong to special solvable classes, such as free-fermion models, stabilizer Hamiltonians, or commuting projector systems~\cite{bravyi2005stabilizer,verstraete2008matrix}. While invaluable for validation and calibration, these models do not adequately probe the capabilities of algorithms intended to address generic interacting quantum systems. Conversely, Hamiltonians that appear generic and strongly interacting typically lack exact reference solutions, making objective benchmarking difficult or impossible.

In this work, we introduce a benchmarking framework that extends the planted-solution paradigm to the setting of general Pauli Hamiltonians, with the explicit goal of enabling controlled tests of quantum ground-state energy estimation algorithms executed on quantum devices. The central idea is to embed an exactly known ground state into a Hamiltonian assembled from frustration-free terms acting on bounded-size blocks, and to expose the resulting model solely through a polynomial-size linear combination of Pauli operators. This Pauli-level representation ensures direct compatibility with quantum algorithms that operate by measuring or estimating expectation values of Pauli observables, such as variational eigensolvers, imaginary-time evolution schemes, and other ground-state energy estimation protocols.

Through a dense Pauli expansion and optional Clifford conjugation~\cite{gottesman1998heisenberg,aaronson2004improved}, the constructed Hamiltonians acquire overlapping supports, noncommuting interactions, and high operator complexity. Crucially, these transformations preserve the spectrum exactly, ensuring that the ground-state energy remains known a priori despite the apparent scrambling of structure. As a result, the framework provides a rigorous and scalable benchmark for assessing the accuracy, robustness, and convergence properties of quantum ground-state energy estimation algorithms on realistic hardware, without relying on classically intractable reference calculations.

The resulting benchmark instances fall into an intermediate regime that is particularly well suited for evaluation. They are efficiently verifiable—exact ground-state energies are known by construction—and the planted structure is not manifest in the Pauli representation presented to the solver. This separation between verifiability and discoverability mirrors the role played by planted solutions in classical complexity theory, providing a principled foundation for benchmarking quantum ground-state algorithms. 
The framework thus provides a controlled and tunable benchmarking environment whose difficulty can range from spectrally benign to conjectured-hard (when classical planted-SAT instances are embedded).
The introduced technique also provides controlled, scalable, and interpretable benchmark instances that can meaningfully stress both classical and quantum solvers. By tuning parameters such as block size, constraint density, and scrambling depth, the framework enables systematic exploration of algorithmic performance across a continuum from weakly structured to strongly scrambled regimes. In doing so, it offers a practical tool for assessing the reliability, scaling behavior, and failure modes of quantum simulation methods in the absence of classical verification.

To support transparency, reproducibility, and broad community use, we make available open-source software implementing the benchmark construction introduced in this work. The software provides an automated pipeline for generating planted-solution Pauli Hamiltonians together with corresponding certification data, including exact ground-state energies. Benchmark instances are produced solely as polynomial-size Pauli expansions, while the planted structure and solution keys are retained separately for validation purposes. All code, documentation, and example instances are freely available in a public GitHub repository~\cite{Hen2026PlantedSolutionBenchmarking}, enabling independent verification of our results and facilitating systematic benchmarking across platforms, algorithms, and hardware architectures (the reader is referred to Appendix~\ref{sec:software} for additional details).

\section{Planted benchmark Hamiltonians}
\label{sec:planted_construction}

\begin{figure*}[t]
\centering
\includegraphics[width=\textwidth]{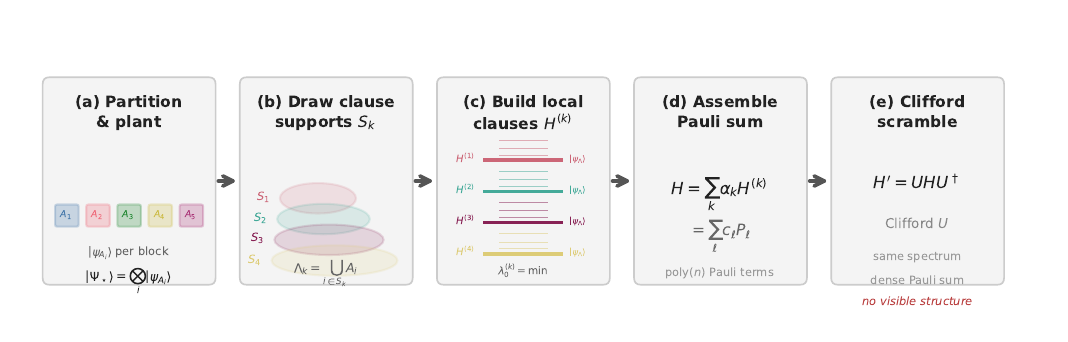}
\caption{Construction pipeline for planted benchmark Hamiltonians. (a)~Qubits are partitioned into disjoint blocks $\{A_i\}$ with planted states $|\psi_{A_i}\rangle$ defining a global product ground state $|\Psi_\star\rangle$. (b)~Random subsets $S_k$ of blocks define overlapping clause supports $\Lambda_k = \bigcup_{i\in S_k} A_i$. (c)~Local clause Hamiltonians $H^{(k)}$ are constructed with the planted restriction as ground state and generic excited-state spectra. (d)~The full Hamiltonian is assembled as a weighted sum $H=\sum_k \alpha_k H^{(k)}$ and expanded in the Pauli basis, yielding a polynomial-size representation with overlapping supports and no visible block structure. (e)~Optional Clifford conjugation $H'=UHU^\dagger$ further scrambles operator support while preserving the spectrum and ground-state energy.}
\label{fig:pipeline}
\end{figure*}

We now introduce a family of benchmark Hamiltonians designed to probe ground-state solvers in a controlled yet nontrivial regime. The construction is guided by four requirements: the Hamiltonians 
should be specified as polynomial-size linear combinations of Pauli strings, and their exact ground states and ground energies should be known by construction, and this ground-state structure should not be manifest in the Pauli representation presented to the solver. Together, these properties ensure that the instances are suitable for systematic benchmarking while avoiding shortcuts to trivial solutions. The overall construction pipeline is summarized in Fig.~\ref{fig:pipeline}.

We consider a system of $n$ qubits partitioned into disjoint blocks
\[
\{A_1, A_2, \dots, A_M\}, \qquad A_i \cap A_j = \emptyset \ (i\neq j),
\]
where each block $A_i$ contains a bounded number of qubits, $|A_i|=O(1)$. The block sizes may be chosen uniformly or at random from a fixed range, and no notion of geometric locality is assumed. For each block $A_i$ we fix a normalized state
\[
|\psi_{A_i}\rangle \in \mathcal{H}_{A_i},
\]
drawn Haar-randomly from the complex unit sphere on $\mathcal{H}_{A_i}$. These block states define a global planted product state
\begin{equation}
|\Psi_\star\rangle = \bigotimes_{i=1}^M |\psi_{A_i}\rangle ,
\label{eq:planted_state}
\end{equation}
which will serve as a ground state of the benchmark Hamiltonians constructed below.

To enforce the planted structure without introducing visible global order, we assemble the Hamiltonian from many locally defined terms with overlapping supports. Specifically, we draw $K$ random subsets
\[
S_k \subseteq \{1,\dots,M\}, \qquad k=1,\dots,K,
\]
according to a fixed distribution, such as uniform sampling over subsets of a given size or independent Bernoulli inclusion of blocks with fixed probability. Each subset $S_k$, of size $|S_k|=O(1)$, determines a support region
\[
\Lambda_k = \bigcup_{i\in S_k} A_i ,
\]
which typically overlaps with those of other terms; in particular, $|\Lambda_k|=O(1)$.

For each region $\Lambda_k$ we define a local Hamiltonian $H^{(k)}$ acting nontrivially only on $\mathcal{H}_{\Lambda_k}$ and extended by identity outside $\Lambda_k$. The operator $H^{(k)}$ is constructed as follows. Let
\[
|\psi_{\Lambda_k}\rangle := \bigotimes_{i\in S_k} |\psi_{A_i}\rangle
\]
denote the restriction of the planted product state to $\Lambda_k$. We choose an orthonormal basis
\[
\{|\phi^{(k)}_r\rangle\}_{r=0}^{d_k-1}, \qquad d_k = 2^{|\Lambda_k|},
\]
of $\mathcal{H}_{\Lambda_k}$ such that $|\phi^{(k)}_0\rangle = |\psi_{\Lambda_k}\rangle$. This basis may be obtained, for example, by completing $|\psi_{\Lambda_k}\rangle$ to a random orthonormal basis via Gram--Schmidt orthogonalization. Without loss of generality, we can take this basis to be separable on the disjoint regions. Since the number of qubits within each block is $O(1)$, the orthogonalization process scales as $O(1)$.

We then assign a real spectrum $\{\lambda^{(k)}_r\}_{r=0}^{d_k-1}$ with
\[
\lambda^{(k)}_0 = \min_r \lambda^{(k)}_r ,
\]
allowing for degeneracy at the minimum if desired, and define
\begin{equation}
H^{(k)} \;=\; \sum_{r=0}^{d_k-1} \lambda^{(k)}_r \, |\phi^{(k)}_r\rangle\langle\phi^{(k)}_r| .
\label{eq:general_local_clause}
\end{equation}
By construction, $|\psi_{\Lambda_k}\rangle$ lies in the ground space of $H^{(k)}$, and hence the global planted state $|\Psi_\star\rangle$ minimizes every local term.

The full benchmark Hamiltonian is defined as a weighted sum of the local clauses,
\begin{equation}
H = \sum_{k=1}^K \alpha_k H^{(k)}, \qquad \alpha_k>0.
\label{eq:big_H}
\end{equation}
The planted state $|\Psi_\star\rangle$ is therefore a ground state of $H$, with exact ground energy
\[
E_0 = \sum_{k=1}^K \alpha_k \lambda^{(k)}_0 ,
\]
which is known explicitly by construction. The ground state of $H$ is unique if and only if the intersection of the local ground spaces of all $H^{(k)}$ is one-dimensional.

Although each $H^{(k)}$ is a completely general Hermitian operator on $\mathcal{H}_{\Lambda_k}$, the total Hamiltonian $H$ remains efficiently representable. Since $|\Lambda_k|=O(1)$, each local clause $H^{(k)}$ admits an exact Pauli expansion with at most $4^{|\Lambda_k|}$ terms. As a result, the full Hamiltonian $H$ can be specified as a polynomial-size linear combination of Pauli strings, despite the local clauses being maximally generic.

It is worth noting that the present framework subsumes classical planted constraint-satisfaction problems as a special case. Specifically, any planted $k$-SAT instance can be compiled into a planted-solution Pauli Hamiltonian within the construction above. Given a planted assignment $z^\star \in \{0,1\}^n$ and a set of clauses $\{C_k\}$ satisfied by $z^\star$, each clause $C_k$ acting on $|S_k| \leq k$ variables maps to a diagonal clause Hamiltonian $H^{(k)}$ whose eigenbasis is the computational basis on the corresponding qubits and whose spectrum assigns eigenvalue $0$ to satisfying assignments and eigenvalue $+1$ to violating ones. Choosing block size $|A_i|=1$ and planted block states $|\psi_{A_i}\rangle = |z^\star_i\rangle$ yields a Hamiltonian $H = \sum_k H^{(k)}$ that is diagonal in the computational basis and whose ground state is $|z^\star\rangle$. This establishes a direct embedding of planted $k$-SAT into the diagonal subclass of the present framework~\cite{jerrum1992large,achlioptas2006solution,barthel2002hiding,krzakala2007gibbs,achlioptas2001two}. Through this diagonal channel, any hardness properties of classical planted constraint satisfaction problems (CSPs) carry over to the quantum construction. However, this hardness transfer applies specifically to instances with computational-basis clause eigenbases; the Haar-random ensemble studied in Sec.~\ref{sec:numerics} exhibits qualitatively different spectral behavior.

To further obscure the planted structure, we optionally apply a global Clifford circuit $U$ of bounded depth,
\begin{equation}
H' = U H U^\dagger .
\label{eq:clifford_conj}
\end{equation}
Crucially, this operation does not alter the planted spectrum. Moreover, because Clifford unitaries map Pauli strings to Pauli strings, $H'$ admits an equally compact Pauli representation, while the planted ground state is transformed into the stabilizer state $U|\Psi_\star\rangle$. We emphasize that this state, although typically highly entangled in the conventional sense, remains efficiently classically describable via the stabilizer formalism~\cite{gottesman1998heisenberg,aaronson2004improved}. Clifford conjugation should therefore be understood as a structural obfuscation step that delocalizes operator support and removes manifest block structure from the Pauli representation, not as a mechanism for producing classically intractable target states. The Hamiltonian $H'$ is the object provided to benchmarking algorithms.

\subsection{Verifiability and discoverability}
\label{sec:hardness}

The construction is designed to occupy an intermediate regime in which the ground state is exactly known to the benchmark designer but not manifest to the solver. Verification is straightforward: given a candidate state $|\Phi\rangle$, its energy with respect to $H' = \sum_\ell c_\ell P_\ell$ can be efficiently estimated by measuring expectation values of Pauli strings, and the planted ground state $U|\Psi_\star\rangle$ achieves the exact minimum energy by design. 

Discovery, in contrast, is substantially more demanding. From the solver's perspective, $H'$ appears as a generic noncommuting sum of Pauli strings with overlapping supports and no explicit tensor-product or commuting-projector structure. Each clause $H^{(k)}$ is a maximally general Hermitian operator on its support, with no commuting or blockwise factorization, and the block partition itself is not provided---inferring it from the Pauli representation amounts to a planted structure recovery problem over an exponentially large space of possibilities. Optional Clifford conjugation further entangles the degrees of freedom, mapping the planted product state to a generally highly entangled one while preserving the compact Pauli-sum form. Although $H$ is frustration free in the hidden basis, this property is invisible in the Pauli representation, where local reductions do not reveal simple spectral projectors.

We emphasize that this separation is not a worst-case hardness claim, and we do not assert that block-structure recovery from $H'$ is computationally hard: the framework establishes no complexity-theoretic hardness and claims no reduction from QMA-hard problems. Indeed, since the planted state is a product state on disjoint blocks, structure-learning approaches based on local correlations or commutation patterns of the Pauli expansion may well recover the partition efficiently in practice, and we have not ruled this out. The construction's purpose is to provide a controlled source of certified reference instances in which the ground-state energy is known by design and the planted structure is not literally read off the Pauli representation; whether a given solver class is meaningfully challenged by these instances is an empirical question that this paper does not address.

A worked example of the entire procedure described above is given in Appendix~\ref{sec:example}.

\section{Numerical characterization of single-planted instances}
\label{sec:numerics}

We now characterize the spectral properties of single-planted instances 
generated by the construction of Sec.~\ref{sec:planted_construction}, 
focusing on how the spectral gap depends on system size~$n$ and clause 
count~$K$. The analysis uses exact diagonalization of instances with 
block size $|A_i|=1$, $|S_k|=3$ blocks per clause, bimodal eigenvalue 
scheme, and clause weights $\alpha_k$ drawn uniformly from $[0.5,1.5]$.
For each pair $(n,K)$ we generate 50 independent instances 
(24--50 for $n=12$) and report medians and interquartile ranges 
across the ensemble. 

The analysis proceeds as follows. We first note that at low clause count, a fraction of instances exhibit numerically degenerate ground states; this regime remains a valid benchmark since the ground energy is known by construction. We then examine how the unnormalized spectral gap grows with $K$, and show that the intensively normalized gap to the first excited state $\Delta_1/K$ grows with clause density $K/n$ over the range studied while closing approximately exponentially in $n$ at fixed density. We then note that the gap profile differs qualitatively from that of classical planted CSPs and trace this to the sum-of-non-negative-penalties form of the construction. Throughout, we report these as spectral features of the ensemble and do not interpret them as statements about the algorithmic difficulty of the instances for any particular solver class. 

\subsection{Ground-space degeneracy at low clause count}

\begin{figure}[t]
\centering
\includegraphics[width=0.9\columnwidth]{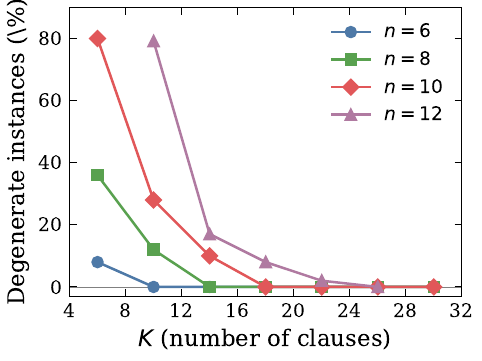}
\caption{Fraction of instances with numerically degenerate ground state ($\Delta_1<10^{-10}$) as a function of $K$ for different system sizes. As one would expect, the degeneracy fraction falls sharply with $K$.}
\label{fig:degeneracy}
\end{figure}

At low clause count, a fraction of instances exhibit numerically degenerate ground states ($\Delta_1 < 10^{-10}$); see Fig.~\ref{fig:degeneracy}. This fraction decreases with both $K$ and $n$, so that for sufficiently large $K$ essentially all instances have a one-dimensional planted ground space (the degenerate regime is not a defect of the construction: the planted state remains an exact ground state of $H$, the ground energy is known by design, and verification of solver outputs against the exact reference value proceeds identically in both regimes).

\subsection{Spectral gap behavior}

We also find that the spectral gap grows 
monotonically and approximately linearly with clause count for all four system sizes studied. This behavior 
follows directly from the sum-of-non-negative-penalties structure of 
the construction: each clause $H^{(k)}$ contributes a term 
$\alpha_k(H^{(k)} - \lambda^{(k)}_0)$ that is non-negative and has the 
planted state in its kernel. Adding clauses can therefore only raise 
excited-state energies while leaving the planted-state energy unchanged, 
making the gap monotonically non-decreasing in $K$.

The linear slope $s(n)$ decreases with system size as
reflecting the fact that each clause constrains a diminishing fraction 
of the system as $n$ grows ($|S_k|/M = 3/n$). 

\begin{figure}[t]
\centering
\includegraphics[width=0.9\columnwidth]{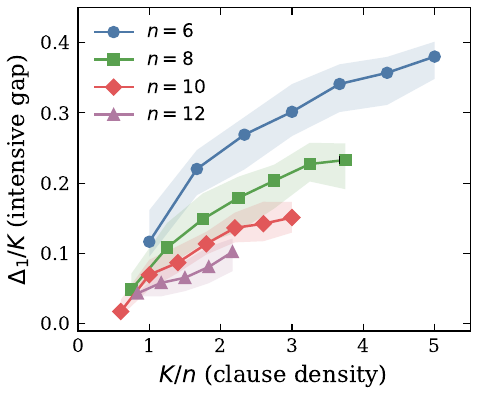}
\caption{Intensive spectral gap $\Delta_1/K$ as a function of clause density $K/n$ for $n=6,8,10,12$. Bands indicate interquartile ranges over 50 instances. At fixed $K/n$, the intensive gap decreases approximately exponentially with $n$.}
\label{fig:intensive_gap}
\end{figure}

If the Hamiltonian is normalized per clause, $H \to H/K$, the intensive spectral gap $\Delta_1/K$ behaves qualitatively differently from its unnormalized counterpart. As shown in Fig.~\ref{fig:intensive_gap}, $\Delta_1/K$ grows with clause density $K/n$ throughout the range studied, without exhibiting clear saturation. More striking is its behavior at fixed clause density: $\Delta_1/K$ decreases approximately exponentially with system size.

Holding the clause count fixed and increasing system size, the 
spectral gap closes exponentially,
\begin{equation}
\Delta_1 \sim c(K) \cdot b(K)^n ,
\label{eq:exp_closing}
\end{equation}
with base $b(K)$ increasing from $\approx 0.70$ at $K=10$ to 
$\approx 0.81$ at $K=26$, and $R^2 > 0.99$ in all cases 
(Table~\ref{tab:exp_closing}). The base increasing with $K$ 
reflects the fact that denser constraint networks slow the 
exponential closing, though they do not prevent it. Equivalently, 
the ``half-life'' of the gap---the number of additional qubits 
required to halve $\Delta_1$---increases from $\approx 1.9$ qubits 
at $K=10$ to $\approx 3.3$ qubits at $K=26$.

\begin{table}[h]
\centering
\caption{Exponential gap closing at fixed $K$. The spectral gap 
follows $\Delta_1 \sim c \cdot b^n$ with $R^2>0.99$ in all cases. 
The half-life $n_{1/2} = -\ln 2 / \ln b$ gives the number of qubits 
required to halve the gap.}
\label{tab:exp_closing}
\begin{tabular}{c c c c}
\hline\hline
$K$ & $b$ & $n_{1/2}$ & $R^2$ \\
\hline
10 & 0.699 & 1.9 & 1.000 \\
14 & 0.750 & 2.4 & 1.000 \\
18 & 0.767 & 2.6 & 0.995 \\
22 & 0.788 & 2.9 & 0.997 \\
26 & 0.810 & 3.3 & 0.995 \\
30 & 0.794 & 3.0 & 0.999 \\
\hline\hline
\end{tabular}
\end{table}

We interpret this exponential closing as an indication that the spectral landscape becomes increasingly compressed with system size, which may challenge certain solver classes, although no claim about algorithmic difficulty is made.

\section{Multiple planted solutions and competing attractors}
\label{sec:m_tuple_planted}

Much like in classical planted SAT, where multi-planted instances are used to defeat solvers that exploit the uniqueness of the planted assignment, the quantum planted-solution framework introduced above admits a natural generalization in which multiple designated global ground states are embedded simultaneously. This multi-planted version offers several advantages over its single-planted counterpart. While the construction continues to yield Hamiltonians with exactly known ground energies and controlled locality, the presence of multiple planted solutions qualitatively alters the structure of the optimization landscape encountered by ground-state solvers. In particular, the energy landscape develops multiple competing global attractors that are locally indistinguishable yet globally incompatible, providing a richer and more demanding class of benchmark instances.

We consider a fixed collection of $m$ global product states
\[
|\Psi^{(a)}_\star\rangle
=
\bigotimes_i |\psi^{(a)}_{A_i}\rangle,
\qquad a=1,\dots,m,
\]
defined on a common block decomposition $\{A_i\}$. For concreteness and robustness, we assume that these states are mutually orthogonal, which may be enforced by choosing the block states such that
\[
\langle \psi^{(a)}_{A_i} \mid \psi^{(b)}_{A_i} \rangle = 0
\qquad \forall i,\ a\neq b.
\]
This blockwise orthogonality ensures that the global planted states remain well separated under local reasoning and small perturbations.

As in the single-planted case, we construct the Hamiltonian from local clauses acting on overlapping regions
\[
\Lambda_k = \bigcup_{i\in S_k} A_i,
\qquad |S_k|=O(1),
\]
with the subsets $S_k$ drawn from a fixed distribution. The key difference lies in the construction of the local clause Hamiltonians. For each region $\Lambda_k$, we define a local Hamiltonian $H^{(k)}$ acting on $\mathcal{H}_{\Lambda_k}$ and extended by identity outside $\Lambda_k$, with the property that its ground space contains the restrictions
\[
|\psi^{(a)}_{\Lambda_k}\rangle
=
\bigotimes_{i\in S_k} |\psi^{(a)}_{A_i}\rangle,
\qquad a=1,\dots,m.
\]

Concretely, for each $\Lambda_k$ we choose an orthonormal basis
\[
\{|\phi^{(k)}_r\rangle\}_{r=0}^{2^{|\Lambda_k|}-1}
\]
of $\mathcal{H}_{\Lambda_k}$ such that
\[
\{|\psi^{(a)}_{\Lambda_k}\rangle\}_{a=1}^m
\subseteq
\mathrm{span}\{|\phi^{(k)}_r\rangle : r \in \mathcal{I}_k\},
\]
for some index set $\mathcal{I}_k$, and assign a real spectrum
\(\{\lambda^{(k)}_r\}\) whose minimum value is attained on all basis states corresponding to $\mathcal{I}_k$. The resulting local clause
\begin{equation}
H^{(k)} =
\sum_r \lambda^{(k)}_r\,|\phi^{(k)}_r\rangle\langle\phi^{(k)}_r|
\end{equation}
is a completely generic Hermitian operator on $\mathcal{H}_{\Lambda_k}$ whose ground space contains all planted local restrictions. No further structure is imposed, and in particular $H^{(k)}$ need not decompose into blockwise or commuting components.

The full Hamiltonian is defined as a weighted sum of these clauses,
\begin{equation}
H = \sum_k \alpha_k H^{(k)}, \qquad \alpha_k>0.
\end{equation}
By construction, every planted global state $|\Psi^{(a)}_\star\rangle$ simultaneously minimizes all local clauses and is therefore a ground state of $H$. The exact ground energy is given by
\[
E_0 = \sum_k \alpha_k \lambda^{(k)}_{\min},
\]
where $\lambda^{(k)}_{\min}$ denotes the minimum eigenvalue of $H^{(k)}$.

The dimension of the global ground space is determined by the intersection of the local ground spaces of all clauses. In contrast to the single-planted case, this intersection is generically $m$-dimensional, with basis vectors given by the planted global states, provided that the overlap structure of the supports $\Lambda_k$ is sufficiently connected. Importantly, this degeneracy is not associated with any explicit symmetry visible in the Pauli representation of the Hamiltonian.

Although each local clause $H^{(k)}$ is a maximally general Hermitian operator on $\mathcal{H}_{\Lambda_k}$, the total Hamiltonian remains efficiently representable. Since $|\Lambda_k|=O(1)$, each clause admits an exact Pauli expansion with at most $4^{|\Lambda_k|}$ terms, and the full Hamiltonian is specified by a polynomial number of Pauli strings.

From an algorithmic perspective, the presence of multiple planted ground states induces a qualitatively different optimization landscape. Each local clause rewards consistency with all planted local restrictions, yet the different planted global states correspond to mutually incompatible global assignments. As a result, a state that partially aligns with one planted configuration on some regions and with a different configuration on others may collect substantial local reward while failing to correspond to any true ground state.

This competition gives rise to multiple basins of attraction that are separated not by explicit frustration or energy barriers, but by global inconsistency between distinct planted explanations. Solvers that rely on local energy gradients may therefore experience slow convergence, oscillatory dynamics, or sensitivity to initialization, drifting between incompatible partial explanations before committing to a particular planted solution.

Finally, as in the single-planted construction, we may apply a global Clifford circuit of bounded depth to conjugate the Hamiltonian without altering its spectrum. The resulting Pauli representation appears as a generic, noncommuting sum of Pauli strings, with no manifest indication of the number, structure, or nature of the planted ground states. These multi-planted instances therefore provide a controlled yet demanding benchmark for probing algorithmic sensitivity to degeneracy, competition, and global structure inference in quantum ground-state solvers.

\section{Conclusions}\label{sec:conclusions}
\label{sec:conclusions}

We have introduced a benchmarking framework for quantum ground-state algorithms based on planted-solution Pauli Hamiltonians and provided a systematic characterization of the resulting instances. The framework constructs Hamiltonians whose ground states and ground energies are known exactly by design, while presenting benchmark instances solely as polynomial-size sums of Pauli operators with no manifest simplifying structure. A key structural feature is that the construction subsumes classical planted constraint-satisfaction problems: classical planted instances can be embedded directly through diagonal clause Hamiltonians, while non-diagonal choices of clause eigenbases allow a continuous interpolation toward genuinely quantum benchmarks.

The spectral behavior of single-planted instances generated with random clause eigenbases is straightforward to summarize. At low clause count, a fraction of instances exhibit degenerate planted ground spaces; these remain legitimate benchmark instances since the ground energy is known by construction. As $K$ grows, the unnormalized spectral gap increases approximately linearly, while the intensively normalized gap $\Delta_1/K$ grows with clause density $K/n$ but closes approximately exponentially with system size at fixed $K/n$. When normalized appropriately, the gap narrows with increasing system size, reflecting the fact that each clause constrains a progressively smaller fraction of the Hilbert space. We report these as spectral features of the ensemble and explicitly refrain from interpreting gap size---or its scaling---as a proxy for the algorithmic difficulty of the instances.

These results characterize the spectral structure of the single-planted random ensemble but do not speak to the algorithmic difficulty of the resulting instances, which would require benchmarking with explicit solvers. The framework also admits two natural extensions beyond the single-planted random ensemble. One is the direct embedding of classical planted instances through the diagonal construction. The other is the introduction of competing multi-planted configurations in which different clauses favor incompatible planted states.

A natural direction for future work is the extension of the planted-solution framework to dynamical benchmarks. In this setting, the goal is to simulate real-time quantum evolution and evaluate nonequilibrium observables while retaining exact reference data. One promising route is to identify subclasses of planted Hamiltonians whose spectra factorize in a hidden representation, enabling efficient classical evaluation of time-dependent quantities. After scrambling, such Hamiltonians would appear as dense noncommuting Pauli sums, while exact dynamical reference values remain available to the benchmark designer. This approach would enable certified dynamical benchmarks in which difficulty arises from representational complexity rather than spectral intractability.

To support reproducibility and community use, we have developed open-source software implementing the benchmark construction. The framework enforces a strict separation between the public benchmark input, provided solely as a Pauli Hamiltonian, and private certification data used for validation. We anticipate that this approach, together with the characterization presented here, will provide a practical benchmark suite for quantum ground-state solvers and a tool for exploring the boundary between classically tractable and quantum-challenging regimes in many-body Hamiltonian optimization.

\section*{Acknowledgments}
The authors would like to thank Lex Kemper for useful comments. This research was sponsored by the Secretary of the Air Force Concepts, Development, and Management (SAF/CDM) organization through the SEQCURE2 program at the University of Maryland's Applied Research Laboratory for Intelligence and Security. 
We acknowledge the Center for Advanced Research Computing (CARC) at the University of Southern California for providing computing resources that have contributed to the research results reported in this study (\url{https://carc.usc.edu}). AI tools were used for stylistic and grammatical editing and for code generation as a programming assistant. All algorithms and specifications were designed by the authors. No AI was used to generate or modify the mathematical results, proofs, numerical experiments, or scientific conclusions. All code was reviewed and validated by the authors.

\bibliographystyle{apsrev4-2}
\bibliography{refs}
\appendix

\section{Software, reproducibility, and instance generation}
\label{sec:software}

To ensure transparency, reproducibility, and broad community use, all benchmark instances used in this work are generated using an open-source software framework. The software implements the planted-solution constructions described in Sec.~\ref{sec:planted_construction} and produces benchmark Hamiltonians together with exact reference data that serve as certification keys for benchmarking purposes.

Benchmark instances are generated through an automated pipeline that mirrors the conceptual structure of the construction. A set of disjoint constant-size blocks $\{A_i\}$ is first defined, and a target block state $|\psi_{A_i}\rangle$ is specified or sampled for each block. A collection of random subsets $\{S_k\}$ is then drawn, determining the supports of the local Hamiltonian terms. From these ingredients, the software assembles a frustration-free Hamiltonian as a sum of local spectral clauses, expands it into a polynomial-size Pauli representation, and optionally applies global Clifford conjugation to scramble operator support while preserving the Pauli-sum form. All sources of randomness are controlled by explicit pseudorandom seeds, ensuring that instances are fully reproducible across platforms and implementations.

For each generated instance, the software produces a corresponding solution key. The solution key contains information that is not provided to benchmarking algorithms, but is used exclusively for validation and analysis. This includes the exact ground-state energy and a compact classical description of the planted ground state. These data enable exact certification of algorithmic outputs without requiring solvers to expose their internal states or intermediate computations.

A central design principle of the framework is the strict separation between benchmark input and validation data. The Hamiltonians distributed for benchmarking are provided exclusively as linear combinations of Pauli strings with real coefficients. No information about the block structure, planted states, random subsets, or Clifford transformations is included in the benchmark input. The solution keys are generated and stored independently and are intended for use only by the benchmark designer or referee. This separation ensures that algorithmic performance reflects genuine solution-finding capability rather than access to hidden structural information. The solution keys should be understood as a certification key for benchmarking, not as a cryptographic secret.

The complete instance-generation code, together with documentation, example benchmark Hamiltonians, and corresponding reference outputs, is made freely available in a public GitHub repository~\cite{Hen2026PlantedSolutionBenchmarking}. Both the software and representative benchmark instances can be downloaded directly and used without modification. This ensures that all results reported in this work are fully reproducible and that the benchmark framework can be readily adopted, extended, and independently validated by the broader community.

\section{Worked example: an explicit small instance} 
\label{sec:example}

We now present a small worked example illustrating the planted block-product construction using fully generic local clause Hamiltonians. The purpose of this example is purely pedagogical: while the benchmark generator employs Haar-random complex block states and generic local spectra, we choose simple, explicit states and spectra here in order to make the construction and resulting Pauli expansions transparent.

\begin{figure}[t]
\centering
\includegraphics[width=\columnwidth]{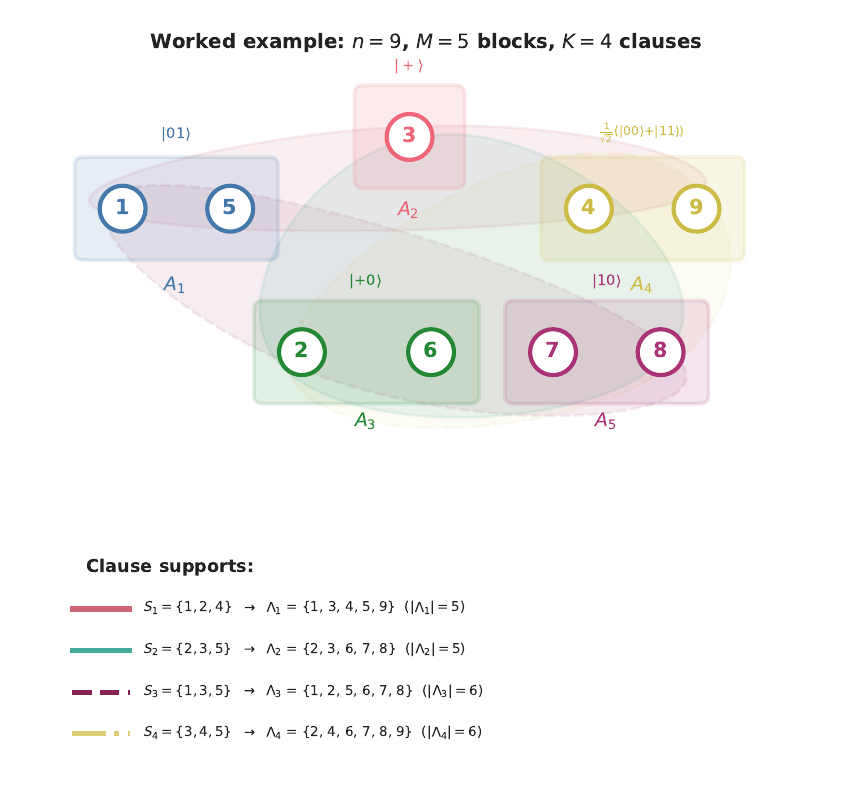}
\caption{Block-clause structure of the worked example. Nine qubits (numbered circles) are partitioned into five blocks $A_1$--$A_5$ (colored backgrounds) with non-contiguous membership. Four clause supports $S_1$--$S_4$ (shaded regions), each spanning three blocks, overlap to produce a fully connected constraint hypergraph. Planted block states are indicated above each cluster.}
\label{fig:worked_example}
\end{figure}

Consider a system of $n=9$ qubits labeled $\{1,2,\dots,9\}$, partitioned into five disjoint blocks of heterogeneous sizes,
\begin{eqnarray}
A_1 &=& \{1,5\}, \quad
A_2 = \{3\}, \quad \\\nonumber 
A_3 &=& \{2,6\}, \quad
A_4 = \{4,9\}, \quad
A_5 = \{7,8\}.
\end{eqnarray}
No geometric locality or ordering is assumed; in particular, each block draws its qubits from non-adjacent positions, illustrating the absence of spatial structure (see Fig.~\ref{fig:worked_example}). For each block $A_i$ we fix a normalized block state
\[
|\psi_{A_1}\rangle = |01\rangle,\quad
|\psi_{A_2}\rangle = |+\rangle,\quad
|\psi_{A_3}\rangle = |{+}0\rangle,
\]
\[
|\psi_{A_4}\rangle = \tfrac{1}{\sqrt{2}}(|00\rangle + |11\rangle),\quad
|\psi_{A_5}\rangle = |10\rangle,
\]
where $|+\rangle=(|0\rangle+|1\rangle)/\sqrt{2}$.
These block states define the planted global product state
\[
|\Psi_\star\rangle
=
\bigotimes_{i=1}^5 |\psi_{A_i}\rangle ,
\]
with qubits ordered according to the block definitions above.

We draw $K=4$ random subsets of blocks, each of size three,
\begin{eqnarray}
S_1 &=& \{1,2,4\}, \quad
S_2 = \{2,3,5\}, \nonumber\\
S_3 &=& \{1,3,5\}, \quad
S_4 = \{3,4,5\},
\end{eqnarray}
which define the corresponding support regions
\[
\Lambda_k = \bigcup_{i\in S_k} A_i .
\]
In this example, $\Lambda_1 = \{1,3,4,5,9\}$ with $|\Lambda_1|=5$, $\Lambda_2 = \{2,3,6,7,8\}$ with $|\Lambda_2|=5$, $\Lambda_3 = \{1,2,5,6,7,8\}$ with $|\Lambda_3|=6$, and $\Lambda_4 = \{2,4,6,7,8,9\}$ with $|\Lambda_4|=6$. The overlap structure ensures that every block participates in at least two clauses: $A_3$ and $A_5$ each appear in three of the four clauses, while $A_1$, $A_2$, and $A_4$ each appear in two. The clause hypergraph is fully connected, with every pair of clauses sharing at least one block.

For each region $\Lambda_k$ we construct a local Hamiltonian $H^{(k)}$ acting nontrivially only on $\mathcal{H}_{\Lambda_k}$ and extended by identity outside $\Lambda_k$. Let
\[
|\psi_{\Lambda_k}\rangle = \bigotimes_{i\in S_k} |\psi_{A_i}\rangle
\]
denote the restriction of the planted state to $\Lambda_k$.

We choose an orthonormal basis
\[
\{|\phi^{(k)}_r\rangle\}_{r=0}^{2^{|\Lambda_k|}-1}
\]
of $\mathcal{H}_{\Lambda_k}$ such that
\[
|\phi^{(k)}_0\rangle = |\psi_{\Lambda_k}\rangle .
\]
For pedagogical simplicity, this basis may be obtained by completing
$|\psi_{\Lambda_k}\rangle$ to a computational basis via Gram--Schmidt orthogonalization, although the generator uses a random orthonormal completion.

We then assign a real spectrum $\{\lambda^{(k)}_r\}$ satisfying
\[
\lambda^{(k)}_0 = \min_r \lambda^{(k)}_r ,
\]
allowing for degeneracy at the minimum if desired, and define
\begin{equation}
H^{(k)} = \sum_{r=0}^{2^{|\Lambda_k|}-1}
\lambda^{(k)}_r\,|\phi^{(k)}_r\rangle\langle\phi^{(k)}_r| .
\label{eq:example_generic_clause}
\end{equation}
By construction, the planted restriction $|\psi_{\Lambda_k}\rangle$ lies in the ground space of $H^{(k)}$.

The full Hamiltonian is defined as a weighted sum of the local clauses,
\begin{equation}
H = \sum_{k=1}^4 \alpha_k H^{(k)},
\end{equation}
with $\alpha_k>0$ (for simplicity, one may take $\alpha_k=1$). The planted global state $|\Psi_\star\rangle$ simultaneously minimizes every clause and is therefore a ground state of $H$. The ground energy is given exactly by
\[
E_0 = \sum_{k=1}^4 \alpha_k \lambda^{(k)}_0 ,
\]
which is known explicitly by construction. The ground state is unique provided that the intersection of the local ground spaces of the four clauses is one-dimensional, which holds generically due to the overlapping support structure.

Although each local clause $H^{(k)}$ is a completely general Hermitian operator on $\mathcal{H}_{\Lambda_k}$, it admits an exact Pauli expansion involving at most $4^{|\Lambda_k|}$ terms. Since $|\Lambda_k|=O(1)$, the full Hamiltonian $H$ is specified by a polynomial number of Pauli strings, even though the local clauses are maximally generic and genuinely entangling across blocks. The resulting Pauli representation exhibits overlapping, non-consecutive supports and lacks a visible block structure. 

Finally, conjugating the Hamiltonian by a global Clifford circuit $U$,
\[
H' = U H U^\dagger ,
\]
preserves the spectrum while further delocalizing operator support across qubits. The resulting Hamiltonian $H'$ appears as a dense, noncommuting Pauli sum with no manifest planted structure, even though its ground energy and planted ground state $U|\Psi_\star\rangle$ remain exactly known.

This worked example illustrates explicitly how a planted block-product ground state can be embedded into a Hamiltonian constructed from generic local clauses with overlapping supports, while remaining efficiently representable as a polynomial-size Pauli expansion.

\end{document}